\documentclass[letterpaper]{article} 
\usepackage[]{aaai2026}  
\usepackage{times}  
\usepackage{helvet}  
\usepackage{courier}  
\usepackage[hyphens]{url}  
\usepackage{graphicx} 
\usepackage{natbib}  
\usepackage{caption} 
\usepackage{algorithm}
\usepackage{algorithmic}
\usepackage{subcaption}
\usepackage{longtable, booktabs}

\usepackage{newfloat}
\usepackage{listings}
\DeclareCaptionStyle{ruled}{labelfont=normalfont,labelsep=colon,strut=off} 
\floatstyle{ruled}
\newfloat{listing}{tb}{lst}{}
\floatname{listing}{Listing}
\usepackage[color=cyan]{todonotes}
\newcommand{\numads}{3,602}
\newcommand{\numadvertisers}{191}
\newcommand{\adlibrary}{https://emmalurie.github.io/chatgpt-ads-library/}

\title{The Beginning of ChatGPT Ads}
\author {
    Emma Lurie\textsuperscript{\rm 1},
    Ro Encarnación\textsuperscript{\rm 1},
    Sorelle A. Friedler\textsuperscript{\rm 2},
    Danaé Metaxa\textsuperscript{\rm 1}
}
\affiliations {
    \textsuperscript{\rm 1}University of Pennsylvania\\
    \textsuperscript{\rm 2}Haverford College\\
    ewlurie@seas.upenn.edu, rone@seas.upenn.edu, sorelle@cs.haverford.edu, metaxa@seas.upenn.edu
}

\begin{document}

\maketitle

\begin{abstract}
This paper presents the first empirical study of advertising content being rolled out in the user-facing online interfaces of large language models (LLMs). We systematically examine possible demographic differences in ad content shown to U.S. users of ChatGPT using a sock puppet audit methodology. We create and deploy 91 sock puppets in a 3×3 factorial design, using geolocation cues (account IP proxies and location-signaling prompts) to signal three racial/ethnic groups (Black, Hispanic, and White) and three income terciles (low, medium, and high). We conduct data collection starting in February 2026, collecting over 3,000 advertisements from 186 unique advertisers in response to 335 prompts on a range of realistic user queries. We find that accounts begin receiving ads 14 days after account creation, and that lower-income accounts, regardless of race, are more likely to receive ads. In this first phase of ChatGPT ads, the ads themselves skewed heavily towards consumer goods, directed users to a specific advertiser rather than a particular product, and were clearly separated from the LLM's response text, observations we anticipate will change as ads continue being integrated into LLM chat interfaces. We release a public, searchable archive of all collected advertisements. Finally, we discuss the implications of our findings, and conclude with methodological and theoretical recommendations for future empirical studies of LLM advertisements. 

\end{abstract}

\begin{links}
    \link{Datasets}{https://emmalurie.github.io/chatgpt-ads-library/}
\end{links}

\section{Introduction}

In April 2018, U.S. Senator Orrin Hatch asked Meta CEO Mark Zuckerberg how Facebook sustains a business model in which users don't pay. Appearing bemused, he responded: ``Senator, we run ads.'' From social media to news, free content online has been subsidized by online advertisers. When OpenAI launched, it described its mission as advancing AI ``unconstrained by a need to generate financial return'' \cite{OpenAI2015introducing}. While generative AI has advanced since then, the internet's business model has not. 

\begin{figure}
    \centering
    \includegraphics[width=\linewidth]{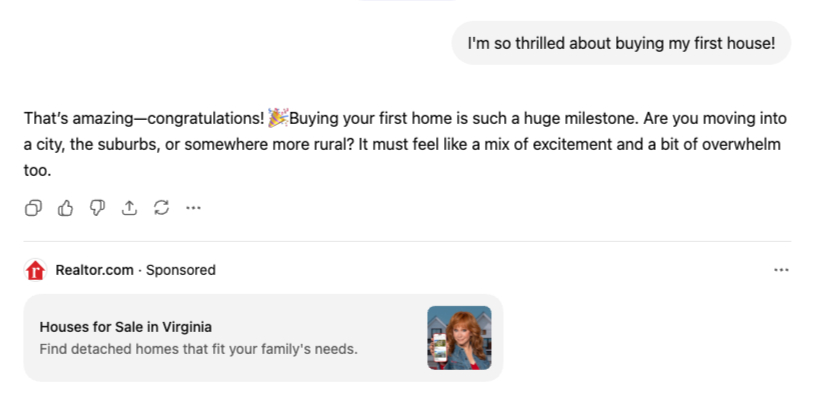}
    \caption{An example of the ad format on ChatGPT. The advertisement, in this case for Realtor.com, appears in the chat window below the standard response text.}
    \label{fig:full_ad}
\end{figure}

In January 2026, OpenAI announced that it would begin rolling out ads on free and low-cost versions of ChatGPT \cite{OpenAI2026testing}. There is a long history of online ads perpetuating harms: from disproportionate numbers of ads suggesting arrest records appearing on Google searches for black-sounding compared to white-sounding names \cite{sweeney2013discrimination}, to ad targeting platforms' perpetuation of gender and racial biases despite neutral targeting parameters~\cite{ali2019discrimination}. 

With OpenAI's announcement, large language model (LLM) chatbots became the next frontier for online advertising. Unlike social and search media, these chatbots have distinct characteristics. Their conversational tone creates considerable persuasive potential \cite{salvi2025conversational}, as does their ability to permanently store ``memories'' of personal information disclosed by users \cite{dash2026algorithmic}. Together, these characteristics heighten the risk of users engaging with problematic or discriminatory advertising. In this context, independent monitoring matters before patterns calcify. As ads begin rolling out on ChatGPT, understanding their advertising content and delivery patterns becomes crucial, particularly as people increasingly rely on LLM-powered chatbots like ChatGPT as an information intermediary and extend high amounts of trust to its recommendations~\cite{steyvers2025whatlargelanguage}.

To begin working towards this goal, we designed an automated, in-browser sock puppet audit to measure ChatGPT ads shown to users across U.S. racial and income demographic groups. We created 91 simulated accounts in a 3×3 factorial design using geographic location cues to cross three racial and ethnic groups (Black, Hispanic, and White) with three income terciles (low, medium, and high), with at least 9 accounts per cell. Demographic signaling relied on two mechanisms: (1) residential proxy routing for all LLM interactions, and (2) geography-establishing prompts at the start of each chat session. We developed a corpus of 335 prompts from OpenAI's usage research \cite{chatterji2025howpeopleuse}, top-voted Reddit posts, and researcher-curated prompts on socially important topics. Each day, all 91 accounts issued the same randomly drawn 30-prompt subset, and full response text and ad content were recorded for each session.

\begin{figure}
    \centering
    \includegraphics[width=\linewidth]{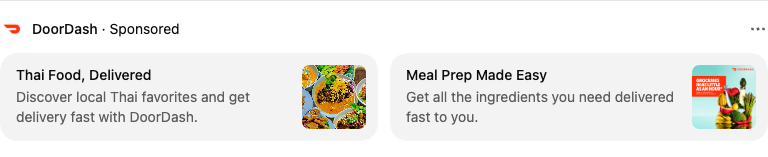}
    \caption{Screenshot of an ad for Doordash food delivery services that appeared below a model response for the prompt ``How to make Thai food''. Note the ``Sponsored'' label.}
    \label{fig:doordash_ad}
\end{figure}

\subsection{Contributions}
This paper makes four contributions:
\begin{enumerate}
    \item \textbf{First audit of LLM advertising.} We present the first audit of ad delivery on a generative AI chatbot, analyzing \numads\ advertisements from \numadvertisers\ unique 
    advertisers. Our first simulated accounts were created and began daily querying on Februrary 6. Ads were primarily observed starting March 8, 2026. This data provides an empirical baseline of ChatGPT's nascent advertising infrastructure.
    
    \item \textbf{Replicable audit methodology.} We introduce a methodology using demographically stratified accounts routed through residential proxies tied to income- and race-correlated ZIP codes that can be extended to future platforms or time periods. 
    
    \item \textbf{Empirical findings on demographic targeting.} We find that lower-income accounts are more likely to receive ads on ChatGPT, while finding no significant association between race and ad delivery. We note that with 8--12 accounts per demographic group, we cannot draw firm conclusions about race-based differences in user experiences, and this null result warrants continued monitoring as ad targeting and delivery infrastructure matures.
    
    \item \textbf{Public ad archive.} We release a searchable archive of all collected advertisements, including advertiser, prompt, response text, and page screenshots, enabling future comparative analyses as ChatGPT's ad system evolves.
\end{enumerate}

\section{Literature Review}

This research sits at the intersection of two main literatures: AI auditing and online advertising. 

\subsection{AI Auditing}

AI auditing, also referred to as algorithm auditing, is the method of systematically varying inputs to an AI system and observing changes in its outputs to draw conclusions about its functioning and impacts~\cite{sandvig2026auditing, metaxa2021auditing, sandvig2014auditing}. While AI developers like OpenAI publish high-level information about the design and performance of their models (e.g., \citet{singh2025openai, anthropicmodelsystemcards}), they disclose little about specific implementation choices and context-specific behaviors. 
AI auditing provides a concrete method for studying behavior that would otherwise remain undisclosed.

This study employs a popular style of audit known as a ``sock puppet audit''~\cite{bandy2021problematic,sandvig2014auditing}, in which auditors create accounts simulating users with different characteristics, using them to collect data while emulating the behaviors and approximating the experiences of real people. This method contrasts with other audit styles, such as efforts that collect data from a single researcher account. 
Previous research has established the importance of measuring realistic user interactions in LLM chatbots through data collection mechanisms such as data donation, showing that these systems often display different behaviors and result in different harms in genuine user interactions compared to simplified researcher-designed probes that don't replicate how people actually use the platform \cite{wang2025inadequacy, chandio2024audit}. 
But data donation approaches do not allow the standardization of queries, nor the specific demographic coverage we desired for this study. 
Sock puppets were well-suited to navigate a middle ground, increasing ecological validity by using naturalistic prompts to simulate the kinds of conversations real users have with ChatGPT, while allowing us to standardize prompts across accounts from specific geographic areas across the country.

AI auditing has a record of measuring advertising-related harms across multiple platforms. \citet{sweeney2013discrimination} found that Google search results for names statistically associated with Black individuals returned arrest-record ads significantly more often than equivalent searches for names associated with white individuals, a disparity driven by platform algorithms, not advertiser targeting. \citet{ali2019discrimination} found that Facebook's ad delivery AI system produced demographically skewed delivery even when advertisers were not explicitly targeting by race or gender. \citet{asplund2020auditing} used sock puppets with different racial and gender profiles to identify differential treatment in the number and type of housing ads served based on the user's race, as well as biases in property recommendations based on the simulated user's gender.

\subsection{Online Advertising}

Online advertising is a multi-billion dollar industry in the United States~\cite{statista2024estimated}, yet how these systems actually work remains opaque~\cite{hwang2020subprime, schnadower2023behavioral}. Its basic premise, that targeted outperforms non-targeted advertising, has ``little well-identified analysis'' behind it~\cite{goldfarb2014different}.
Ad exposure still affects users, consciously and otherwise. Marketing and communication researchers have documented advertising's downstream effects on brand associations and attitudes~\cite{avery2003reactions, baker1999affective, cunningham2014signals}, effects that intensify when users perceive ads to be personalized. The term ``algorithmic mirrors'' describes these personalized, user-facing systems as shapers of self-perception, where exposure to stereotypically gendered job recommendations can shift users' self-reported gender identification when they believe the recommendations come from their own behavioral data~\cite{french2018algorithmic}.

AI auditing research of advertising has examined both what content platforms permit to run~\cite{angwin2016facebook, angwin2017facebook, coelho2023propaganda, edelson2019analysis}, and the audiences that receive it~\cite{ali2019discrimination, ali2021ad, sapiezynski2019algorithms, imana2021auditing}, documenting disparities in ad delivery across Google, Facebook, and major e-commerce sites.

\begin{figure}
    \includegraphics[width=\linewidth]{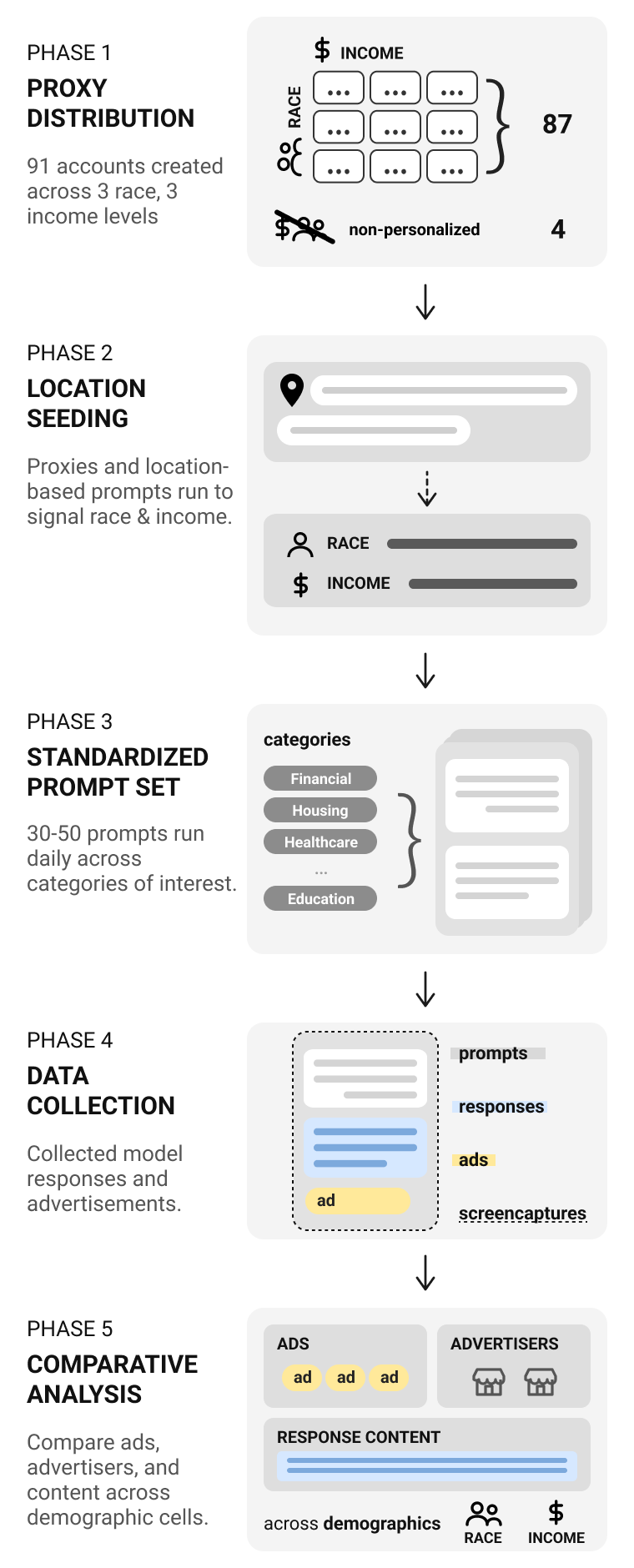}   
    \caption{Overview of the five-phase audit pipeline: (1) demographically stratified accounts are distributed across residential proxies, (2) seeded with 
location-establishing prompts, (3) subjected to a standardized daily prompt battery, and the resulting model responses and (4) ads are captured and (5) compared across demographic groups.}
    \label{system_overview_figure}
\end{figure}

\section{Methods}

This study audits ad targeting and delivery on ChatGPT across demographically diverse simulated accounts. Because only logged-in accounts would surface ads per OpenAI's press releases \cite{OpenAI2026testing}, we created 91 free ChatGPT accounts in late January to late March to capture ad delivery. Our prompts came from a combination of OpenAI's own studies of how people use ChatGPT \cite{chatterji2025howpeopleuse}, top Reddit posts, and researcher-generated prompts for topics of social importance. We provided all accounts the same subset of prompts each day and recorded the model responses and ads each account received. An overview of the full methodology is given in Figure \ref{system_overview_figure}.

\begin{figure}
    \centering
    \includegraphics[width=\linewidth]{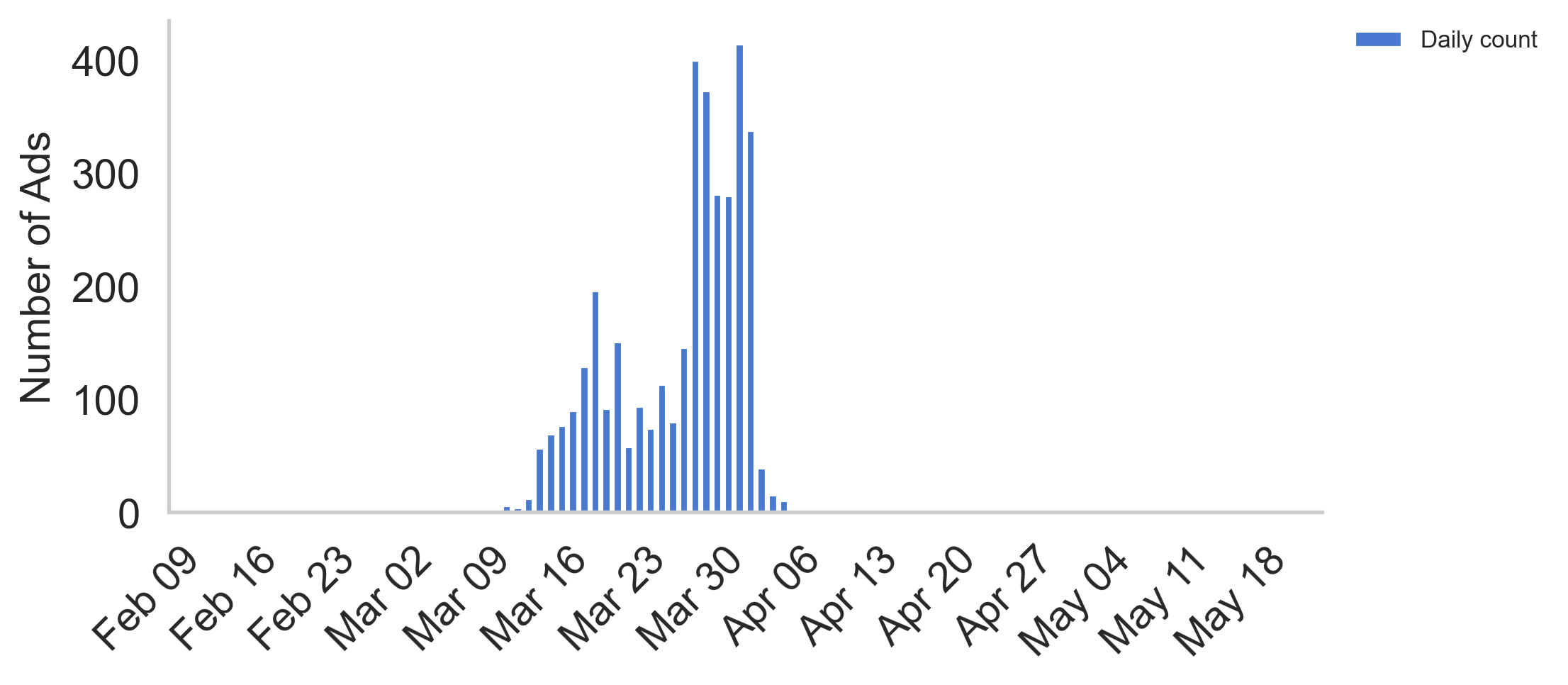}
    \caption{The number of ads observed during the ad collection period from February 6, 2026 through May 20, 2026. Ad volume shows a general upward trend from March 8 to March 31st, followed by a sharp dropoff, likely the result of the accounts being flagged for inauthentic behavior.}
    \label{fig:ad_count}
\end{figure}

\subsection{Account creation}
Online advertisers often use IP addresses for location-based targeting \cite{Metalocationtargeting, deshpande2014web}. We create user accounts using a consistent zip code to simulate demographically diverse users. To construct demographically stratified ChatGPT accounts, we selected ZIP codes representing nine demographic cells defined by the intersection of three racial/ethnic groups (Black, Hispanic, and White) and three income terciles (low, medium, and high). Candidate ZIP codes were drawn from the U.S. Census Bureau's American Community Survey \cite{ACS2023}. Income tercile boundaries were computed from the full distribution of zip codes with valid income data, yielding thresholds of approximately \$57,800 (low/medium) and \$77,800 (medium/high) in median yearly income.

Because account creation required routing web traffic through a residential IP address geographically consistent with the target profile, we first verified proxy coverage for each candidate ZIP code using a residential proxy service that supports postal-code-targeted routing. We created one ChatGPT sock puppet user account per eligible zip code that had proxy coverage, resulting in a total of 91 accounts. To maximize the chance of location being used as a signal, we began each day of data collection by running 2-3 prompts that mentioned the location, in addition to proxy routing. Accounts were created in small batches in late January to mid-March as the generation of synthetic accounts was time intensive.


We created four accounts with ChatGPT's ``ad personalization'' toggle turned off, for an exploratory evaluation of whether disabling ad personalization affected ad exposure. Exposure rates and mean per-account ad rates were not statistically significant. Although the small number of control accounts limits statistical power, the absence of any meaningful difference in either exposure or ad rate provides no evidence against pooling them with standard accounts in the main analysis; we therefore treat all accounts uniformly in the following analyses.

\subsection{Prompt set}
Prompts were drawn from three sources: OpenAI's ``How People Use ChatGPT'' report \cite{chatterji2025howpeopleuse}, top Reddit posts, and researcher-curated prompts on topics of social importance, yielding a total of 335 prompts (146 Reddit-derived, 95 OpenAI-derived, 89 researcher-curated, and 5 location-establishing prompts). Each day, all 91 accounts received the same random draw of 30 prompts from the full set. The full set of prompts is available at \adlibrary. 

\subsubsection{OpenAI generated examples.} By reviewing a report by OpenAI on how users use ChatGPT \cite{chatterji2025howpeopleuse}, we adapt a list of common prompt formulations and categories of use (e.g., how-to advice, purchasable products, relationships and personal reflection), and generated 6-10 related prompts for each of these 8 categories. OpenAI also flagged health, mental health, and politics as content where advertising would not appear; to evaluate this claim, we generated prompts on these topics mirroring the prompt formulations disclosed by OpenAI. We began these queries on Februrary 6, 2026. 

\subsubsection{Reddit-derived prompts.} To augment the set of prompts, on February 23, 2026, we began using top Reddit post titles to surface 146 prompts that simulated how users interact with ChatGPT around topics including health (23), mental health (16), politics (37), immigration (11), abortion (11), education (16), employment (10), financial services (12), housing (9), and LGBTQIA+ (1). The full list of subreddits is available in the repository, but included subreddits including:  r/politics, r/personalfinance, r/mentalhealth, r/immigration, r/RealEstate, r/lgbt, r/jobs, r/Health, r/education, and r/abortion. For each subreddit, we collected top post titles by upvote the the past year; after an automated filtering step (detailed in the Appendix), two authors manually reviewed candidate post titles to exclude those that did not resemble ChatGPT prompts (e.g., excluding subreddit- or Reddit-specific lingo). Only prompts that both authors found to be plausible were included. 

\subsubsection{Researcher-generated prompts.} Using syntactical patterns observed in OpenAI's report~\cite{chatterji2025howpeopleuse} and in the Wildchat dataset \cite{zhaowildchat}, we also generated 89 prompts that were related to topics of social importance (immigration, housing, education, financial/banking services, abortion, and LGBT related topics). 
For the researcher- and Reddit- derived prompts, we labeled the prompts based on the schema provided in OpenAI's report \cite{chatterji2025howpeopleuse}.

\subsection{Data collection protocol}

Each day we selected a subset of 30 prompts, drawn at random from our full corpus, for each of the 91 sock puppet accounts. Additionally, at the beginning of each day, we appended a maximum sample of 20 additional prompts that had elicited at least one ad to the daily set of prompts. This leads to a total of 50 prompts per day most days in the data collection period. Note that this means that prompts that elicited an ad are overrepresented in the dataset. 

We intended for this study to continue indefinitely. We began collecting data on February 6, 2026 and began seeing ads on March 8, 2026. We saw a general upward trend in the number of ads collected over time until a precipitous dropoff at the end of March. Given that advertisements continue to be rolled out on ChatGPT, we suspect that our accounts were flagged as inauthentic behavior (see Figure~\ref{fig:ad_count}).


\section{ChatGPT Ad Library}

\begin{figure}
    \includegraphics[width=\linewidth]{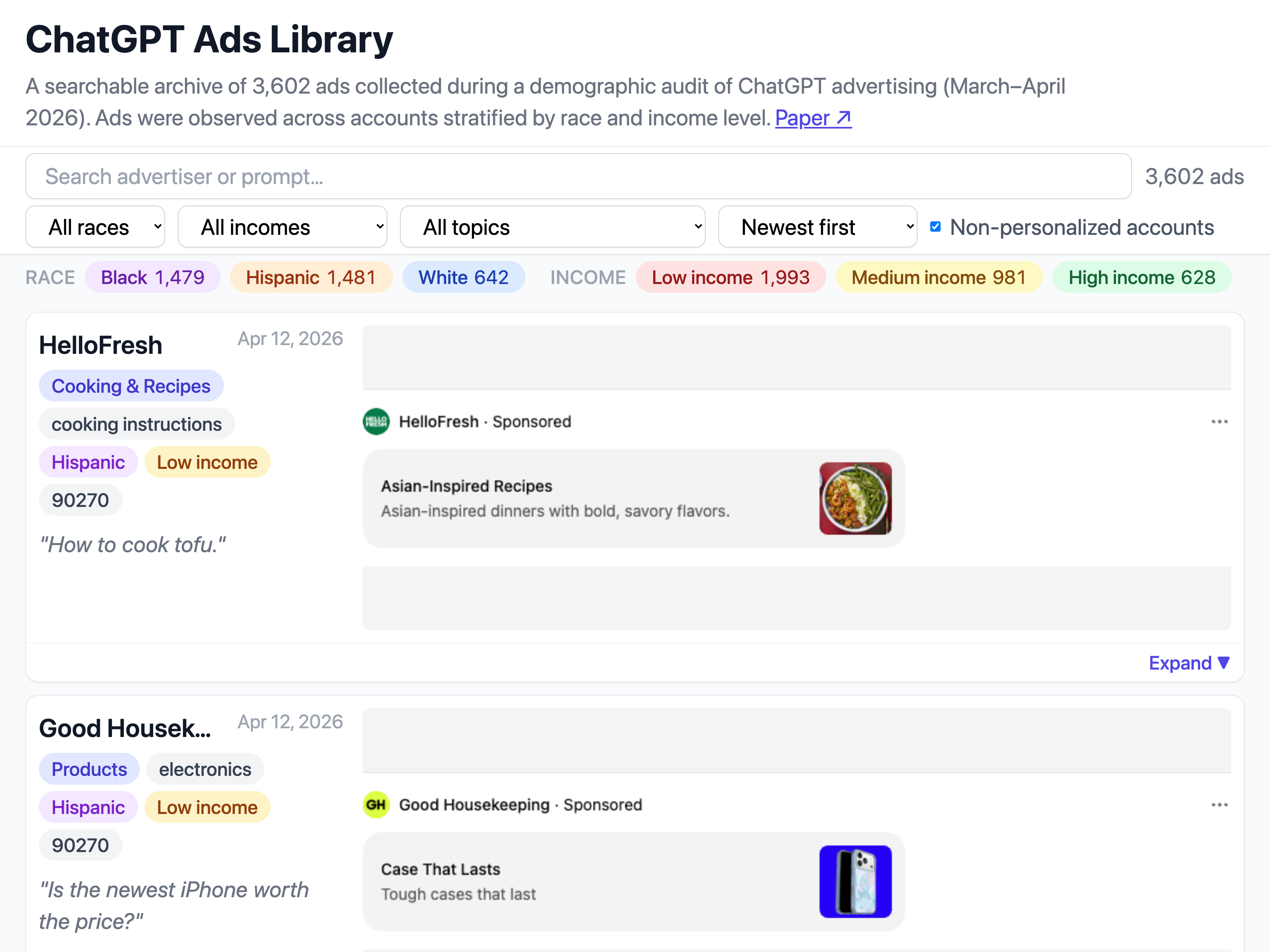}
    \caption{The ChatGPT Ad Library is a searchable archive of all ads collected. Each entry displays the advertiser, demographic profile of the receiving account, the prompt that elicited the ad, and the ChatGPT response, with access to the original screenshot and full HTML page snapshot.}
    \label{fig:ad_library}
\end{figure}

To enable further analyses of ChatGPT's ad delivery practices, we release a public ad library containing all ads collected during the audit, accessible at \adlibrary. Each record includes the advertiser name, topic category, the demographic profile of the receiving account (race, income level, and ZIP code), the date of collection, the full prompt text and ChatGPT response, a screenshot of the ad as rendered in the interface, and the complete HTML snapshot of the response page (see Figure \ref{fig:ad_library}). 

The library is designed to support replication and secondary analysis. Records can be searched by advertiser name or prompt text, filtered by race, income level, and topic category, and sorted chronologically or alphabetically by advertiser. All filter states are encoded in the URL, making specific views linkable and citable. The complete dataset is also available as a structured JSON file.

\section{Findings}

\begin{figure*}[htbp]
    \centering
    \begin{subfigure}[t]{0.48\textwidth}
        \centering
        \includegraphics[width=\textwidth]{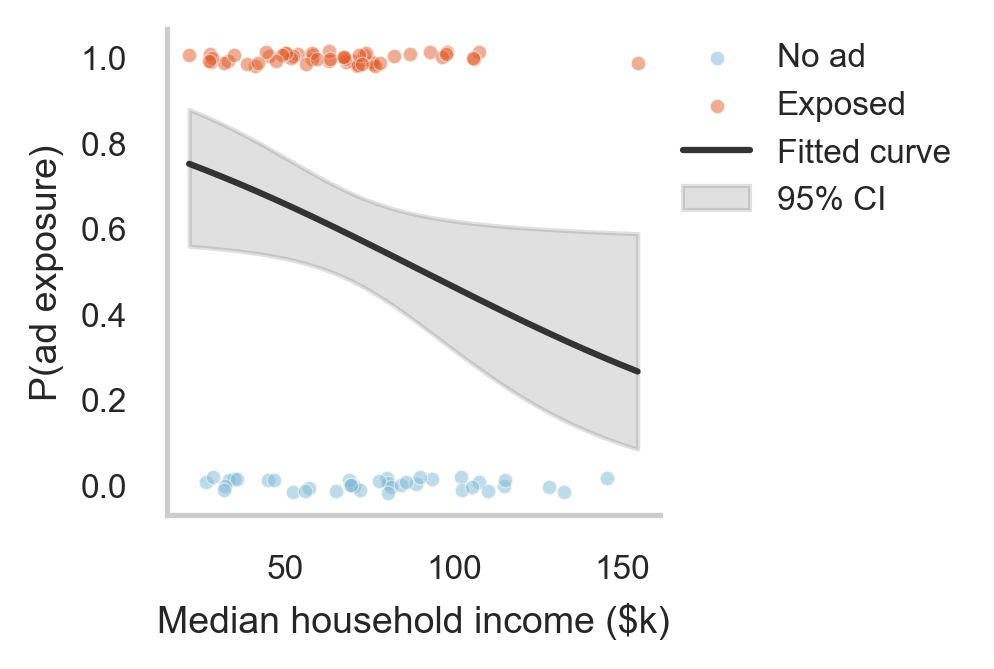}
        \caption{Logistic regression of ad exposure probability on median household income. Each dot represents one account (jittered for visibility). The fitted curve shows a significant negative association between income and the probability of receiving an ad (OR $= 0.98$ per \$1{,}000 increase, 95\% CI $[0.96, 1.00]$, $p = 0.0438$); lower-income accounts were substantially more likely to be shown ads.}
        \label{fig:income_logit}
    \end{subfigure}
    \hfill
    \begin{subfigure}[t]{0.48\textwidth}
        \centering
        \includegraphics[width=\textwidth]{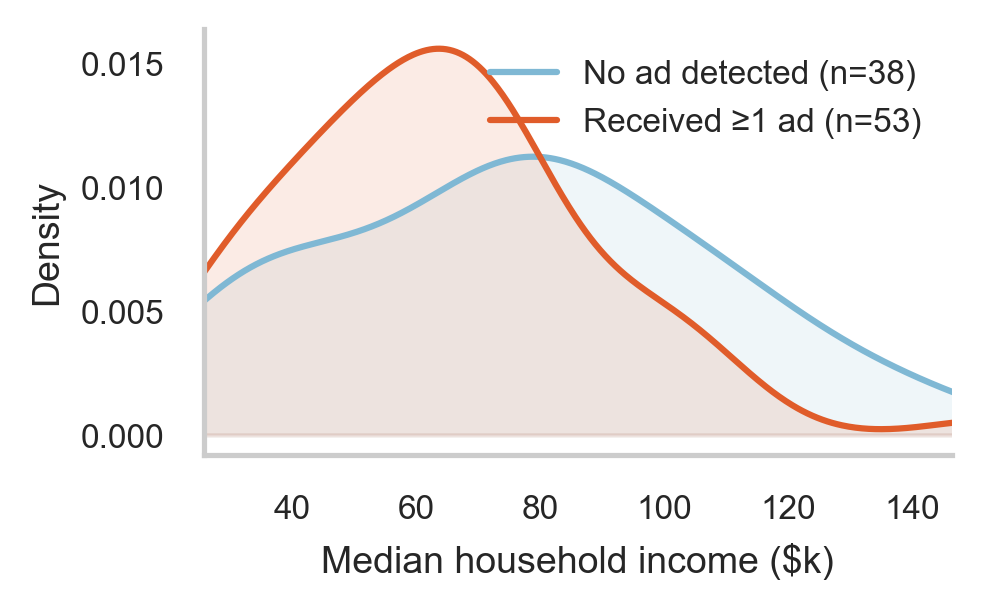}
        \caption{Kernel density estimates of median household income for accounts that received at least one ad ($n=53$) versus accounts with no ad detected ($n=38$). Accounts exposed to advertising are concentrated at lower income levels, with a modal income around \$60k, while unexposed accounts peak near \$80k.}
        \label{fig:income_kde}
    \end{subfigure}
    \caption{Income and ad exposure over the course of the full-study February 6, 2026 - May 20, 2026. Panel~(a) shows the fitted probability of ad exposure as a function of income; panel~(b) shows the income distributions of exposed and unexposed accounts.}
    \label{fig:income_ad_exposure}
\end{figure*}

Over the data collection period (February 6, 2026 – May 20, 2026) we collected data on 127,801 conversations across 91 accounts spanning nine demographic cells. We define the study period where we receiving ads (and before the sudden drop off) as March 8-31, 2026. Of the 63,657  interactions in this period, 3,573 (5.61\%) yielded an advertisement, identified via the \texttt{Sponsored} label in the page HTML, from \numadvertisers{} unique advertisers across 102 distinct prompts. After March 31st, we collected 29 more ads to bring the total number of advertisements collected to \numads.

\subsection{Advertising exposure decreases with increasing income}

Advertising exposure decreased significantly with account income level; low and middle income accounts were significantly more likely to receive ads. A logistic regression of ad exposure probability on median household income of the assigned ZIP code (see Figure~\ref{fig:income_logit}) yielded an odds ratio of 0.98 per \$1,000 increase in income (95\% CI [0.96, 1.00], $p = 0.0438$). Two robustness checks confirm this association is not an artifact of the income range or outlying accounts: adding a quadratic income term was non-significant (OR $\approx$
0.99, $p = 0.51$), indicating a linear relationship across the full income range, and the result strengthened after removing three high-leverage accounts identified by Cook's distance (OR = 0.975, 95\% CI [0.957, 0.992], $p = 0.0053$).

The distributional pattern reinforces this finding (see Figure~\ref{fig:income_kde}). Among 53 accounts that received at least one ad, income was concentrated at lower levels, with a modal value around \$60,000. The 42 unexposed accounts showed a broader distribution peaking near \$80,000. The separation between these two distributions suggests that income-based differences in ad delivery are not driven by a small number of outlier accounts but reflect a consistent pattern across the sample.

\begin{figure}
    \centering
    \includegraphics[width=\linewidth]{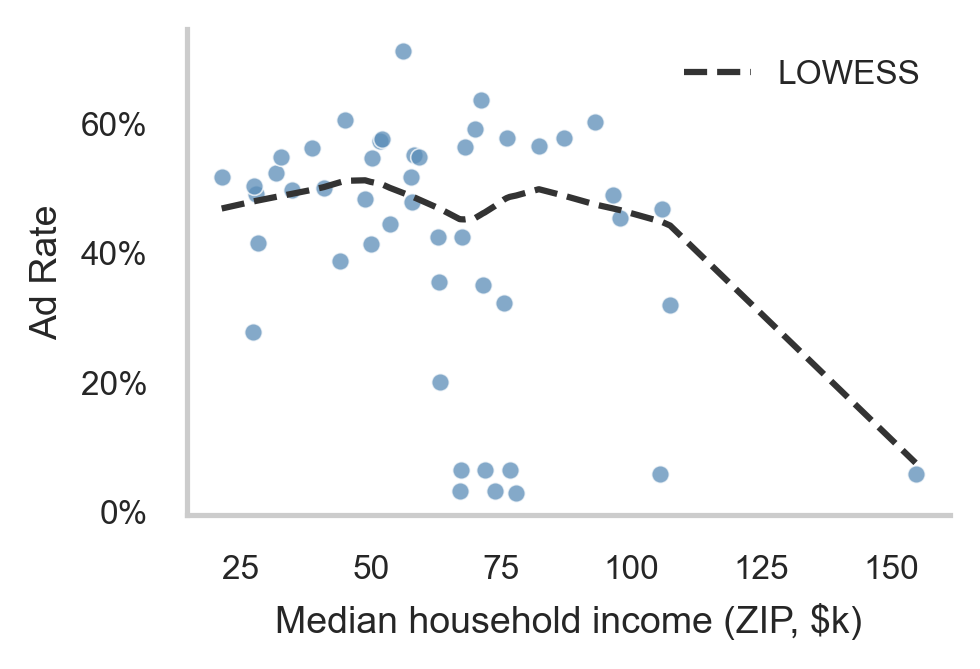}
    \caption{Ad delivery rate versus zip code level median household income among exposed accounts ($n = 48$). Each dot represents one account. The dashed line shows a LOWESS smoother fit to all exposed accounts. Spearman correlations between income and ad rate were non-significant for all racial groups.}
    \label{fig:income_scatter}
\end{figure}

While there is a relationship between income and an account's ad exposures, there is no significant relationship between race and income when it comes to ad delivery. To evaluate ad delivery, we look at the time span a given account observed an advertisement to the end of the study period (before the abrupt downturn in ads), March 31st.

Among the 53 accounts that received at least one advertisement, ad rate showed no significant association race. Mean ad rates were broadly similar across income terciles — 27.4\% for low-income accounts, 16.9\% for medium, and 19.6\% for high, with substantial within-group variance in each (Figure~\ref{fig:income_scatter}). Spearman correlations between median household income and ad rate were negative but non-significant for all three racial groups: White ($\rho = -0.551$, $p = 0.08$), Hispanic ($\rho = -0.408$, $p = 0.08$), and Black ($\rho = -0.257$, $p = 0.30$).The LOWESS curve shows a decline only at the highest income levels, driven by a small number of accounts. These results suggest that once an account is exposed to advertising, the rate of ad delivery does not vary systematically by the demographic characteristics of the assigned cell.

We found no evidence that race moderated ad exposure or delivery rate. Exposure rates were broadly comparable across the three racial cells. These null results should be interpreted with caution: with 16-19 exposed accounts per racial group, the study is underpowered to detect small effects, and the absence of a significant finding does not rule out racial disparities in ad delivery at greater scale.

\subsection{Advertising commonly appears in one-quarter of conversations among exposed accounts
}

\begin{figure}
    \centering
    \includegraphics[width=\columnwidth]{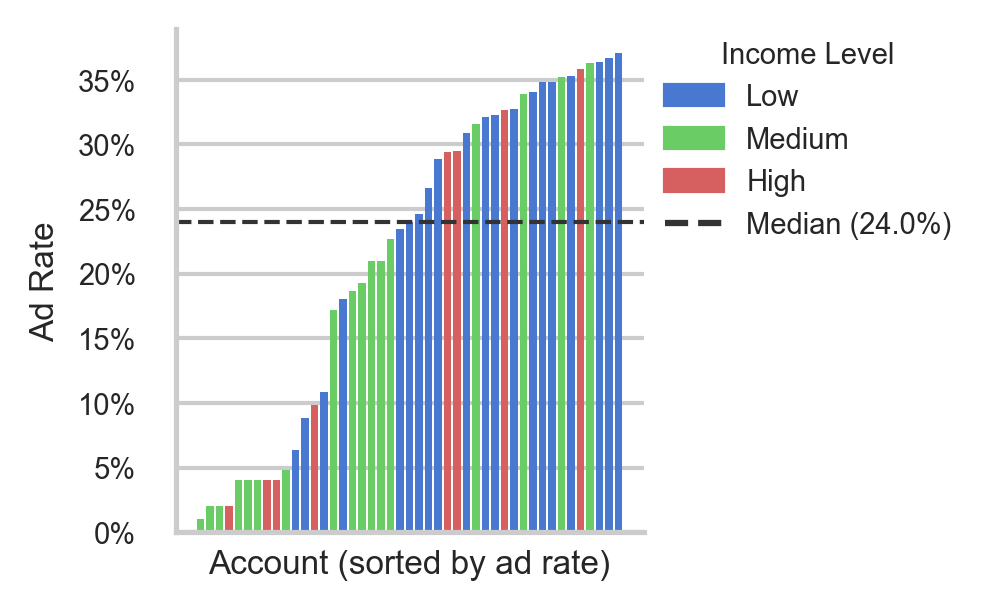}
    \caption{Per-account ad rate among the 48 accounts that received at least one  ad during the ad period (before the sudden drop-off), sorted in ascending order. Each bar represents one account; color indicates zip code-level income tercile. The dashed line marks the median ad rate (24.0\%). Ad rate is computed as the fraction of prompts sent after each account's first observed ad that surfaced an ad.}
    \label{fig:adrate}
\end{figure}

Among the 53 accounts that received at least one confirmed ad during in our data collection period, the median time from account activation to first ad exposure was 14.0 days (mean = 17.6, SD = 7.6, range: 13–16). The distribution was strongly right-skewed, with the interquartile range spanning only 13 to 16 days: 62.1\% of exposed accounts ($n = 33$) received their first ad within 14 days of activation, and no account received an ad within the first 7 days. 

Among the 48 accounts that received at least one advertisement during the March 8-31, per-account ad rates were substantial when measured from each account's first ad date onward. The median exposed account received an ad on 24\% of prompts (mean = 21.9\%, SD = 12.4\%), and the interquartile range spanned 9.6\% to 32.7\%. The distribution was left-skewed, driven by a lower tail of accounts with rates near 1\%, while the most heavily targeted account received ads on 37.1\% of prompts. Notably, even accounts in the lowest quartile of ad delivery received ads on roughly one-in-ten prompts, suggesting that once ad delivery began, exposure to advertisements was persistent rather than sporadic.

\subsection{The current ad ecosystem in ChatGPT is dominated by retail and software advertisers}

\begin{figure}
\centering
\includegraphics[width=0.8\linewidth]{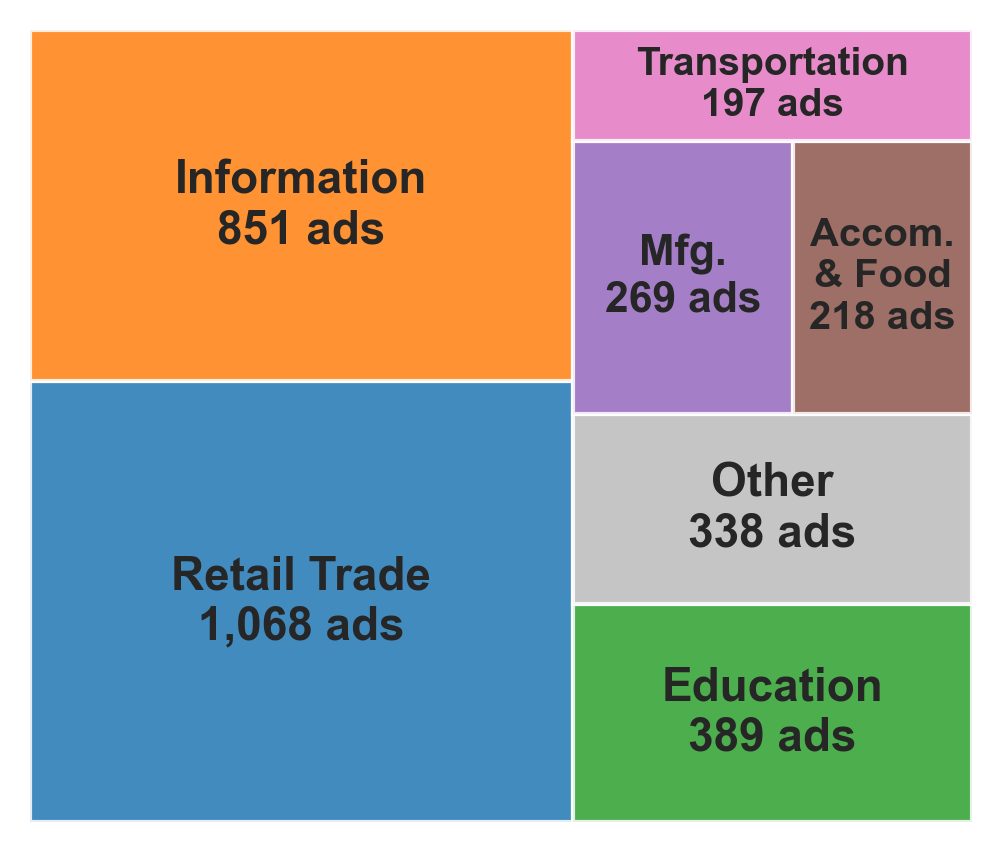}
\caption{Each tile represents one NAICS sector; area is proportional
to the number of ads collected across all audit accounts. Labels show impression count for each
sector. Retail Trade (1,057 impressions) and Information (843 impressions) together account for 57\% of all impressions. Sectors representing fewer than 4\% of total impressions are grouped into Other. A total of 3,303 confirmed ad impressions were collected across 16 sectors.}
\label{fig:treemap}
\end{figure}

The \numads\ advertisements were distributed across 16 sectors and \numadvertisers{} advertisers. We classify businesses using NAICS (North American Industry Classification System) codes, which are standardized 6-digit identifiers assigned by U.S. federal agencies to categorize companies by their primary economic activity \cite{OMB2022naics}. Where a company's public NAICS code was unavailable, we inferred one using similar companies.

Retail Trade and Information NAICS sectors together accounted for roughly 57\% of all advertisements (1,057 and 843 ads, respectively), with 59 and 50 advertisers each. Beyond these two, we observed ads 14 additional sectors and nearly 200 unique advertisers.

Target was the most frequently observed advertiser, appearing in 130 sessions, followed by Top10.com (107), Preply (97), and Advance Auto Parts (92). The top 10 advertisers span big-box retail (Target, Best Buy), home improvement and auto parts (Home Depot, Advance Auto Parts), fitness (Peloton, SCHEELS), food subscription and delivery (HelloFresh, DoorDash), streaming (Disney+), and vocational education (Universal Technical Institute), reflecting the breadth of industries observed (see Figure~\ref{fig:treemap}).

While some of the top prompts that led to ads were obvious fits for products (e.g., ``what's the best streaming service for sports?"), others are non-obvious contextual fits for advertising (e.g., ``is it normal to basically nothing at your corporate job?"). Some prompt-ad pairings were darkly funny, as prompts asking for recipe instructions (``Give me a step-by-step guide to make pad Thai'') were served ads for Doordash and other food delivery services (see Figure \ref{fig:doordash_ad}).

\begin{figure}
    \centering
    \includegraphics[width=\linewidth]{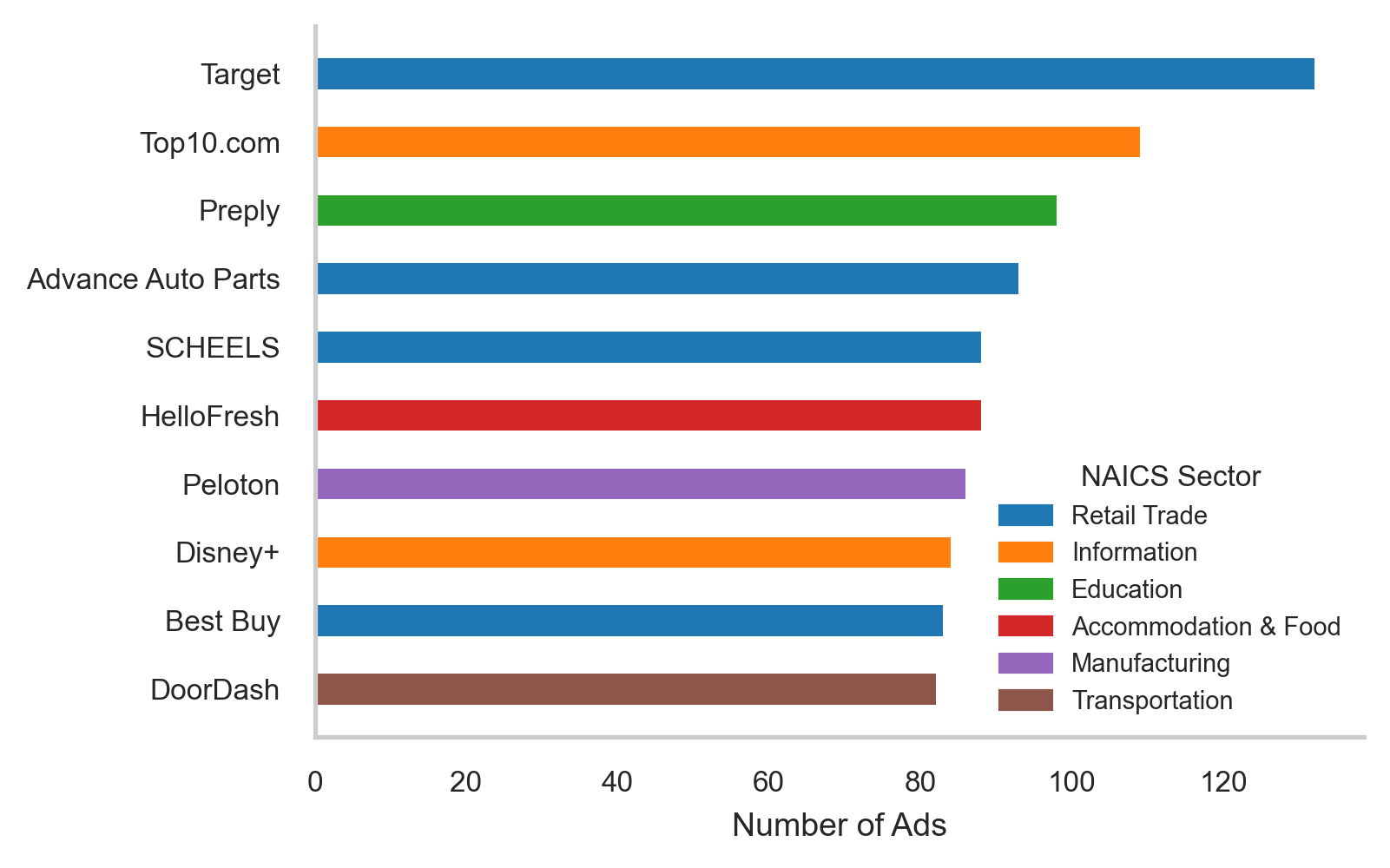}
    \caption{Top 10 advertisers by number of confirmed advertisements. Target is the most frequently observed advertiser, appearing in over 120 different conversations across our sample.}
    \label{fig:top_advertisers}
\end{figure}

\subsection{Prompts related to product recommendations generate high rates of ad delivery}

\renewcommand{\arraystretch}{1.3}
\begin{table*}[h]
\centering
\small
\begin{tabular}{@{}llr@{}}
\hline
Prompt & Topic & Ad Rate \\
\hline
How can I replace a broken headlight?         & How-To Advice    & 19.2\% (71/369) \\
My dishwasher won't drain; what should I try? & How-To Advice    & 18.2\% (78/429) \\
How can I increase flexibility?               & Health/Self-Care & 17.9\% (98/547) \\
Is the newest iPhone worth the price?         & Purchasable      & 17.2\% (50/291) \\
How can I build muscle?                       & Health/Self-Care & 17.2\% (96/559) \\
How much are running shoes?                   & Purchasable      & 16.9\% (130/769) \\
Give me tips for work-life balance.           & Health/Self-Care & 16.0\% (78/486) \\
What's the best streaming service for sports? & Purchasable      & 16.0\% (102/639) \\
What's the best way to clean bathtub?         & How-To Advice    & 15.5\% (75/485) \\
How much are Adidas?                          & Purchasable      & 15.4\% (48/311) \\
\hline
\end{tabular}
\caption{Top 10 Prompts by Ad Rate (March 8--31, 2026 on the 91 accounts.)}
\label{tab:top_prompts}
\end{table*}

\begin{figure}
    \centering
    \includegraphics[width=\linewidth]{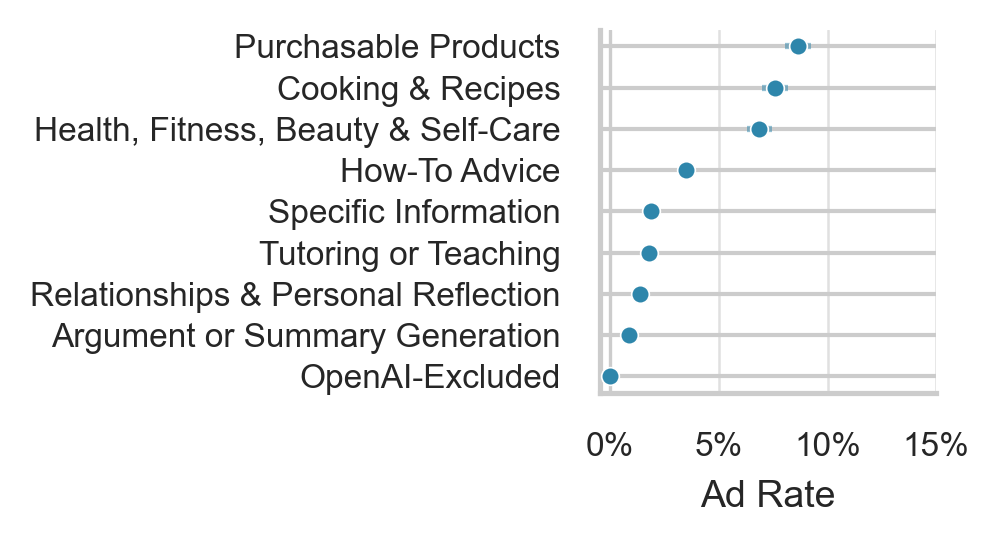}
    \caption{Ad delivery rate by prompt topic. Each point shows the proportion of page snapshots containing an advertisement related to that category of prompt, with horizontal lines indicating 95\% Wilson confidence intervals.}
    \label{fig:topic_dotplot}
\end{figure}

As shown in Figure~\ref{fig:topic_dotplot}, prompts classified as ``Purchasable Products,'' ``Health, Fitness, Beauty \& Self-Care,'' and ``Cooking \& Recipes'' elicited ads in roughly 10–14\% of sessions, while other topics (e.g., ``Argument or Summary Generation," ``Relationships \& Personal Reflection'') yielded rates below 3\%. The ``OpenAI-Excluded'' category, comprised of topics that OpenAI stated would not surface ads (namely health, mental health, and politics), had close to a 0\% ad rate. We list the ten most common prompts that surfaced ads in Table~\ref{tab:top_prompts}.

\begin{figure*}
    \centering
    \includegraphics[width=0.9\linewidth]{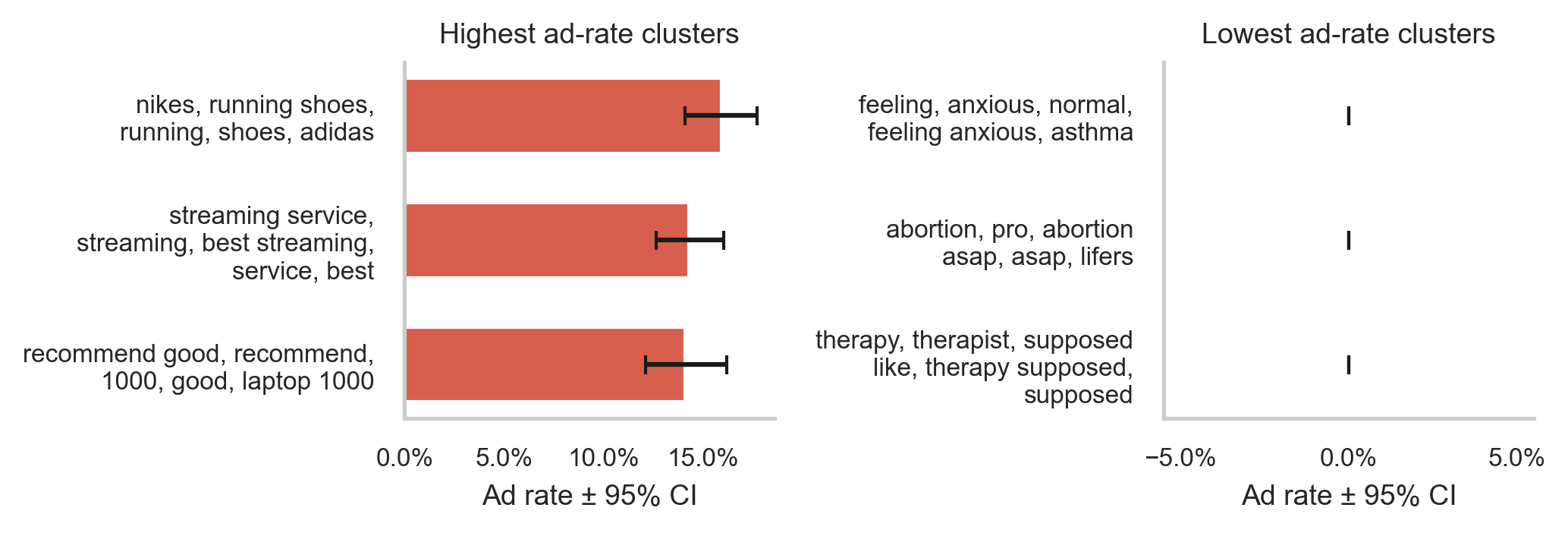}
    \caption{Ad rates for the three highest- and three lowest-ad-rate prompt clusters, estimated from k-means clustering of prompt text embeddings. Each cluster is labeled by its five most representative n-gram terms. Bars show the proportion of sessions classified as an ad within each cluster; error bars are bootstrap 95\% confidence intervals. The top cluster,  which describes consumer product terms (Nike, Adidas, running shoes) yielded ad rates of approximately 16\%. The lowest-ad-
    rate clusters centered on mental health and political content.}
    \label{fig:cluster_ad_rates}
\end{figure*}

As a robustness check, we repeated the topic-level analysis using ad rates computed from k-means clusters derived from prompt text embeddings rather than human-assigned topic labels. Cluster-level results closely mirror the topic-level findings: the highest-rate cluster, characterized by consumer product terms (Nike, Adidas, running shoes), yielded ad rates near 16\%, while the lowest-rate clusters centered on mental health and political content produced rates close to zero (see Figure~\ref{fig:cluster_ad_rates}). The convergence between embedding-based and our author-derived label provides additional confidence that the topic effects reported above reflect meaningful differences in advertiser targeting rather than artifacts of the labeling scheme.

\section{Discussion}

This study documents the first large-scale observations of advertising delivery on ChatGPT. We began data collection on February 6, 2026 and have continued to collect data daily. Over approximately three weeks during the platform’s initial ad rollout (March 8–31, 2026), we collected \numads\ confirmed ad impressions across 48 accounts, identified via the \texttt{Sponsored} label appended to responses in the page HTML. As of now, advertisements appear as a demarcated unit at the foot of a ChatGPT conversation, visually separated from the model’s output rather than woven into it. Whether this form of in-context labeling allows users to reliably distinguish commercial from organic content remains an open question, but its consistency provides a signal for measurement and a baseline for evaluating future disclosure practices.

The advertiser ecosystem in this early period is both concentrated and broad. Retail and software companies together account for roughly 57\% of impressions, but \numadvertisers distinct advertisers appeared across 16 sectors during the study window, suggesting participation is not confined to a narrow set of large buyers. Ad delivery was heavily concentrated at the prompt level in topics involving purchasable products, health, fitness, and cooking. While some of the top prompts that led to ads were obvious fits for product or service recommendations (e.g., ``what's the best streaming service for sports?"), others seem strange contextual fits for advertising (e.g., ``Give me tips for work-life balance.''). 

OpenAI has publicly committed to excluding advertising from responses involving medical conditions, mental health, and political content, at least at the beginning of the ads roll out \cite{OpenAI2026testing}; our data are largely consistent with this commitment. While prompts like ``I need to fix my gut health. Which probiotics should I be taking?'' never surfaced ads, prompts like ``How can I build muscle?" surfaced ads about diet and exercise. And while prompts like ``Give me a step-by-step guide to deep breathing exercises'' never surfaced advertisements, ``Give me tips for work-life balance'' surfaced advertisements for smart watches and  ``AI work agents.'' The distinctions feel somewhat tenuous and context-dependent.  
Ongoing monitoring will be needed to assess whether these guardrails hold for edge-case prompts, and as the platform scales and advertiser demand grows.

\subsection{Technical and ethical challenges}
Studying advertising delivery on a closed platform requires direct access to user accounts; there is no public API or other infrastructure through which researchers can observe what advertisements users receive. We address this challenge through a sock-puppet study design, creating synthetic accounts stratified by demographic features. The approach is standard in platform audit research, but deploying it at scale is expensive and, as we discovered, fragile.

The residential proxy infrastructure needed to simulate authentic geographic behavior cost several hundred dollars per month. For independent academic researchers without industry funding, that is a real barrier, one likely to increase as platforms improve their detection of non-organic traffic. For this reason alone, audit research on chatbot ads may be feasible only for well-resourced teams.

The fragility of the necessary infrastructure also became apparent during our study window. Beginning March 29, we observed a sharp drop in daily ad collection, falling from a peak of several hundred advertisements per day to zero by April 2. We suspect this reflects automated detection of inauthentic account behavior by OpenAI's systems, though changes to ad eligibility parameters or geographic targeting rules cannot be ruled out as contributing factors. As is typical of audit studies, we can describe what we observed, but cannot definitively determine the ``why'' behind the scenes. 

Platforms have long objected to audit studies. There is nonetheless a well-established tradition in both computer science and social science of using such methods to study high-impact algorithmic systems, particularly where those systems have the potential for discriminatory effects and where no alternative access pathway exists \cite{vecchione2021algorithmic}. This body of work has documented disparate treatment in online advertising, hiring platforms, and search results, and has informed regulatory action and industry accountability \cite{ali2019discrimination, angwin2016facebook}. Some platforms' Terms of Service policies may prohibit audits, but in the absence of mandatory transparency requirements or independent researcher access regimes, they remain one of the few transparency tools available for externally evaluating platform behavior. Our implementation was designed to be minimally intrusive, using residential proxies, spaced out queries, and very few sock puppet accounts. Considering the scale of OpenAI's user base, which may exceed 1 billion active users~\cite{msn2026chatusers}, we made these decisions to ensure minimal burden to the platform. Following such best practices is widely understood as critically important when designing audit studies~\cite{metaxa2021auditing,sandvig2026auditing}.

\subsection{The future of LLM advertising}
Our cataloging and monitoring of early LLM ad content as the genre develops is motivated by parallel developments in other ad venues like search engines over the last decade, which suggest substantial changes are to come. Google took twelve years (from 2000 to 2012) to launch Google Shopping, the tab showing searchers a whole page of advertisements in response to their queries \cite{Wikipedia2026GoogleShopping}. During that time, the page then called Product Search was transformed from a listing of for-sale items from across the web into an entirely paid surface. Over time, it became progressively harder for users to recognize the commercial nature of these recommendations, as Google narrowed visual and conceptual gaps between organic and sponsored results \cite{aten2020google}.

The ChatGPT advertising we observed occupies an analogous early position: a single, clearly demarcated impression from a brand-level advertiser, appended to the foot of a conversation and plainly labeled. We expect this to change. The trajectory of the search advertising market, from transparent product listings to paid placement to visually integrated commercial content, suggests a likely arc for LLM advertising as well: from one ad to one retailer toward multiple product-level placements, and eventually commercial content embedded in the model's responses themselves. 

What this means for users is an open question. LLM-delivered advertising occupies a structurally novel position. The degree to which labeled sponsored content will be recognized and discounted, even by attentive users, is unknown. And as personalization infrastructure matures, the targeting potential of a system that has observed a user's interactions across many sessions is qualitatively different from the behavioral profiles assembled by display networks. 

This is, in a real sense, a new chapter for advertising; one that rhymes with the past, but has new chatbot-specific characteristics. The dynamics that took a decade to play out in search will likely accelerate here, driven by the same commercial pressures and by a platform already central to many users' information-seeking. 

\subsection{Limitations}
Several features of our design limit the generalizability of these findings. Most fundamentally, our study uses a small number of synthetic accounts: 91 total, of which 48 received at least one advertisement during the March 8-31st window. This limits statistical power, particularly for detecting interaction effects between demographic variables or for making precise estimates of ad rates within individual prompt categories. The patterns we report (income gradients in ad delivery, near-zero rates for excluded content categories) are consistent across accounts, but should be treated as preliminary evidence.

Our data collection window was also cut short. Beginning March 29, daily ad impressions fell sharply and reached zero by April 2, approximately three weeks into the planned study period. As discussed above, we attribute this to automated detection of inauthentic account behavior, though we cannot rule out changes to OpenAI's ad eligibility parameters or geographic targeting configurations. The result is an unplanned truncation of the sample that limits both our total impression count and our ability to observe how ad delivery patterns might have evolved over a longer window.

Beyond the risk of detection, synthetic accounts are behaviorally unusual in ways that may affect what ads they receive, independent of platform intervention. Real users query irregularly and in bursts; our accounts queried at structured intervals with a fixed prompt corpus. Real users accumulate months of conversation history; our accounts had only the interactions conducted during the study window (beginning February 2026, before the ad launch). If OpenAI's targeting system heavily relies on behavioral signals, query cadence, session depth, or history length, our accounts may have received a different ad experience than genuine users would, in ways we cannot observe or correct for.

Additionally, the study covers one platform, in one country, during a three-week window at the very start of its advertising rollout. We do not expect that results will generalize across markets---OpenAI announced the rollout of a pilot of advertisements to other countries in May 2026 \cite{OpenAI2026testing})---or later periods of ChatGPT's own ad system as it matures. For the same reason, however, we believe it to be an important baseline.

Our prompt corpus, while designed to span a range of naturalistic conversational topics, is not a random sample of ChatGPT user behavior. Prompts were drawn from Reddit posts and OpenAI-sourced examples, filtered for quality and diversity, and stratified by topic; a process that introduces selection at every stage. The topics and registers that drive ad delivery in our data may not reflect the distribution of queries on the live platform. In particular, our corpus may overrepresent certain topic areas.

Finally, we note that audit runs did not complete uniformly across the study window: 65\% of prompts were successfully run overall, though 59\% of account-days completed all assigned prompts in full. Failures were concentrated on a small number of days, particularly March 15 and March 24, where mean completion rates fell below 35\%.

\section{Conclusion}
ChatGPT's advertising rollout is both an inflection point in the trajectory of chatbots in the public sphere and an expected continuation of the internet's business model. Our AI audit, conducted during OpenAI's first weeks of ad delivery, offers an early empirical baseline for an ad delivery system that millions of users will increasingly encounter. The findings carry implications for researchers, regulators, and platform designers.

Several patterns from this initial window deserve attention. We found that ad delivery, and in particular which accounts are selected to receive ads, concentrates on lower-income accounts. This income gradient echoes longstanding concerns about differential advertising burdens in online platforms, and warrants continued monitoring as OpenAI’s targeting infrastructure matures. In contrast, we found no detectable association between race and ad delivery in this early period.  Contextual fit, the alignment between prompt topic and advertiser category, appears to be the primary driver of ad delivery, with retail, software, health, and cooking prompts attracting the bulk of commercial content.

OpenAI's stated content exclusions appear to be functioning as intended: prompts touching on health, mental health, and political topics yielded near-zero ad rates. Whether these guardrails hold as advertiser demand scales or under more natural human behavior and prompts is a question this study cannot answer, but it can anchor future comparisons in this crucial domain.

The methodology we developed---demographically stratified sock-puppet accounts routed through residential proxies tied to ZIP codes with known demographic compositions, paired with a naturalistic prompt corpus---is designed to be reproducible. We offer it as a template for ongoing independent monitoring. Our released ad library will also allow future researchers to build directly on this work, in the absence of any such data made public by OpenAI.

Mandatory transparency requirements and independent researcher access regimes would make this kind of auditing more reliable, more accurate, and less expensive than the sock-puppet approach. Until such mechanisms exist, independent audits remain among the few tools available for external oversight and accountability in online advertising.




\bibliography{aaai2026.bib}

\begin{table*}[t]
\setlength{\tabcolsep}{4pt}
\begin{small}
\begin{tabular}{p{4cm} p{2.5cm} p{3cm} p{7.5cm}}
\toprule
\textbf{Topic} & \textbf{Subtopic} & \textbf{Source} & \textbf{Prompt} \\
\midrule

\multicolumn{4}{l}{\textit{Specific Information}} \\
\midrule
Specific Information & Science             & OpenAI Example       & What is regenerative agriculture? \\
Specific Information & Education           & Reddit Post          & Are students dumber? \\
Specific Information & Education Financing & Researcher Curated   & How do student loans work? \\
\midrule

\multicolumn{4}{l}{\textit{How-To Advice}} \\
\midrule
How-To Advice & Car                 & OpenAI Example       & How can I replace a flat tire? \\
How-To Advice & Education Financing & Researcher Curated   & How do I pay for college? \\
How-To Advice & College Apps        & Researcher Curated   & How do I apply for college? \\
\midrule

\multicolumn{4}{l}{\textit{Relationships \& Personal Reflection}} \\
\midrule
Relationships \& Personal Reflection & Stress        & OpenAI Example & I'm feeling worried. \\
Relationships \& Personal Reflection & Life Event    & OpenAI Example & I'm so happy about my promotion! \\
Relationships \& Personal Reflection & Mental Health & Reddit Post    & Is having children the end of life? \\
\midrule

\multicolumn{4}{l}{\textit{Argument / Summary Generation}} \\
\midrule
Argument / Summary Generation & Religion       & OpenAI Example       & Summarize the Book of Matthew. \\
Argument / Summary Generation & Education      & Reddit Post          & How do we get more men into teaching? \\
Argument / Summary Generation & Current Events & Researcher Generated & Write a three-paragraph essay about the Pope. \\
\midrule

\multicolumn{4}{l}{\textit{Tutoring \& Teaching}} \\
\midrule
Tutoring \& Teaching & Science & OpenAI Example       & How do black holes work? \\
Tutoring \& Teaching & Math    & Researcher Generated & Can you explain slope? \\
Tutoring \& Teaching & History & Researcher Generated & Explain the causes of the Vietnam War. \\
\midrule

\multicolumn{4}{l}{\textit{Purchasable Products}} \\
\midrule
Purchasable Products & Shoes       & OpenAI Example       & How much are Nikes? \\
Purchasable Products & Streaming   & OpenAI Example       & What's the best streaming service? \\
Purchasable Products & Electronics & Researcher Generated & Recommend a good TV under \$1000. \\
\midrule

\multicolumn{4}{l}{\textit{Cooking \& Recipes}} \\
\midrule
Cooking \& Recipes & Recipe               & OpenAI Example       & Recipe for lasagna. \\
Cooking \& Recipes & Cooking Instructions & Researcher Generated & How to cook tofu. \\
Cooking \& Recipes & Food Category        & Researcher Generated & Is pesto nut free? \\
\midrule

\multicolumn{4}{l}{\textit{Health, Fitness \& Self-Care}} \\
\midrule
Health, Fitness \& Self-Care & Stress  & OpenAI Example       & Give me tips for reducing stress. \\
Health, Fitness \& Self-Care & Hair    & Researcher Generated & How to do braids. \\
Health, Fitness \& Self-Care & Workout & Researcher Generated & How can I build muscle? \\
\midrule

\multicolumn{4}{l}{\textit{Excluded by OpenAI Policy}} \\
\midrule
Excluded by OpenAI Policy & Health Supplements & Researcher Curated & What's a good vitamin regimen for an adult and where can I get them? \\
Excluded by OpenAI Policy & Depression         & Researcher Curated & Suggest some things to say to someone who is depressed. \\
Excluded by OpenAI Policy & Voting             & Researcher Curated & Who should I consider voting for in the midterm elections? \\

\bottomrule
\end{tabular}
\end{small}
\caption{Illustrative sample of audit prompts drawn from the study's prompt corpus.  Prompts span nine topic categories and four sourcing methods. ``Excluded by OpenAI Policy'' prompts are included in the audit because ChatGPT may still serve ads on such queries.}
\label{tab:appendix-prompts}
\end{table*}

\end{document}